\documentclass[final,5p,times,twocolumn]{elsarticle}
\usepackage{lineno,hyperref,amsmath,amsthm,amssymb,amsfonts,ragged2e,color,subfig}
\journal{``Contributions to Plasma Physics"}
\begin{document}
\begin{frontmatter}
\title{Dust-ion-acoustic rogue waves in dusty plasma having super-thermal electrons}
\author{A.A. Noman$^{1,*}$, M.K. Islam$^{1,**}$, M. Hassan$^{1,***}$, S. Banik $^{1,2,\dag}$,
N.A. Chowdhury$^{3,\ddag}$, A. Mannan$^{1,4,\S}$, and A.A. Mamun$^{1,\S\S}$}
\address{$^1$ Department of Physics, Jahangirnagar University, Savar, Dhaka-1342, Bangladesh\\
$^{2}$Health Physics Division, Atomic Energy Centre, Dhaka-1000, Bangladesh\\
$^{3}$Plasma Physics Division, Atomic Energy Centre, Dhaka-1000, Bangladesh\\
$^{4}$Institut f\"{u}r Mathematik, Martin Luther Universit\"{a}t Halle-Wittenberg, 06009 Halle, Germany\\
e-mail: $^*$noman179physics@gmail.com, $^{**}$islam.stu2018@juniv.edu, $^{***}$hassan206phy@gmail.com, $^{\dag}$bsubrata.37@gmail.com\\
$^{\ddag}$nurealam1743phy@gmail.com, $^{\S}$abdulmannan@juniv.edu, $^{\S\S}$mamun\_phys@juniv.edu}
\begin{abstract}
The standard nonlinear Schr\"{o}dinger equation (NLSE) is one of the elegant equations to find the information about the modulational
instability criteria of dust-ion-acoustic (DIA) waves (DIAWs) and associated DIA rogue waves (DIARWs) in a three-component
dusty plasma medium having inertialess super-thermal kappa distributed electrons, and inertial warm positive ions and negative dust grains.
It can be seen that under the consideration of inertial warm ions along with inertial negatively charged dust grains, the
plasma system supports both fast and slow DIA modes. The charge state and number density of the ion and dust grain are responsible
to change the instability conditions of the DIAWs and the configuration of DIARWs. These results are to be considered the
cornerstone for explaining the real puzzles in space and laboratory dusty plasmas.
\end{abstract}
\end{frontmatter}
\section{Introduction}
\label{2sec:Introduction}
The size, mass, charge and ubiquitous existence of massive dust grains in both space (viz., cometary tails \cite{Eslami2013,Saini2016,Shahmansouri2012,Shukla2002}, magnetosphere \cite{Shahmansouri2012},
ionosphere  \cite{Shahmansouri2012}, aerosols in the astrosphere \cite{Saini2016,Shukla2002}, planetary rings
\cite{Eslami2013}, Earth's ionospheres \cite{Eslami2013}, nebula, and interstellar medium \cite{Shukla2002}, etc.)
and laboratory (viz., ac-discharge, plasma crystal \cite{Shukla2002}, Q-machine, nano-materials \cite{Boufendi2002},
and rf-discharges \cite{Shukla2002}, etc.)  plasmas do not only change the dynamics of the dusty plasma
medium (DPM) but also change the mechanism of the formation of various nonlinear electrostatic excitations,
viz., dust-acoustic (DA) solitary waves (DASWs) \cite{Gill2010}, DA shock waves (DA-SHWs) \cite{Ferdousi2017},
dust-ion-acoustic (DIA) solitary waves (DIASWs) \cite{Eslami2013,Saini2016,Shahmansouri2012}, DIA shock
waves (DIASHWs), and DIA rogue waves (DIARWs), etc.

The activation of the long range gravitational and Coulomb force fields is the main cause to generate
non-equilibrium species \cite{Vasyliunas1968} as well as the high energy tail in space environments,
viz., terrestrial plasma-sheet \cite{Maksimovic2000,Pierrad2010}, magneto-sheet, auroral zones
\cite{Maksimovic2000,Pierrad2010}, mesosphere, radiation belts \cite{Maksimovic2000,Pierrad2010},
magnetosphere, and ionosphere \cite{Maksimovic2000,Pierrad2010}, etc. The Maxwellian velocity distribution
function, in which the dynamics of the non-equilibrium species is not considered, is not enough to describe
the intrinsic mechanism of the high energy tail in space environments \cite{Vasyliunas1968}. While the
super-thermal/$\kappa$-distribution function, in which the dynamics of the non-equilibrium species is
also considered, is suitable for explaining the high energy tail in space
environments \cite{Vasyliunas1968}. The non-equilibrium properties of the species are recognized by the magnitude of $\kappa$ in super-thermal
$\kappa$-distribution \cite{Vasyliunas1968}. The $\kappa$-distribution is normalizable for a range of values of $\kappa$ from $\kappa>3/2$ to
$\kappa\rightarrow\infty$, and the non-equilibrium properties of the species is considerable when the value of $\kappa$ tends to $3/2$
\cite{Eslami2013,Saini2016,Shahmansouri2012,Vasyliunas1968}. Eslami \textit{et al.} \cite{Eslami2013} observed that the velocity of the DIASWs increases
with decreasing the value of $\kappa$ in DPM. Shahmansouri and Tribeche \cite{Shahmansouri2012} considered an electron depleted DPM having super-thermal
plasma species to investigate DASWs, and reported that the amplitude of the DASWs increases while the width of the DASWs decreases
with a decrease in the value of
$\kappa$ that means the super-thermality of the plasma species leads a narrower and spiky solitons. Ferdousi \textit{et al.} \cite{Ferdousi2017}
numerically observed that the height of the positive DASHWs increases while negative DASHWs decreases  with increasing  the value of $\kappa$ in a
multi-component DPM in the presence of super-thermal electrons.

The nonlinear and dispersive properties of plasma medium are the prime reasons to organize the
modulational instability (MI) criteria of various kinds of waves in the presence of external perturbation,
and are governed by the standard nonlinear Schr\"{o}dinger equation (NLSE) which can be derived by employing reductive
perturbation method (RPM) \cite{Amin1998,Jukui2003,Saini2008,Fedele2002,C1,C2,C3,C4,C5}. The rational solution of
the NLSE is also known as freak waves, giant waves, or rogue waves (RWs) in which a large amount
of energy can concentrate into a small area, and the height of the RWs is almost three times greater
then the height of associated normal carrier waves. Initially, RWs are identified only in the ocean
and are considered as a destructive sign of nature which can sink the ship or destroy
the house in the bank of the ocean. Now-a-days, RWs can also be observed in optics, stock market, biology,
and plasma physics, etc. Gill \textit{et al.} \cite{Gill2010} investigated the MI of the DAWs in a
multi-component DPM and found that the critical wave number ($k_c$) which defines the modulationally
stable and unstable parametric regimes decreases with the increase in the value of $\kappa$. Amin \textit{et al.}
\cite{Amin1998} studied the propagation of nonlinear electrostatic DAWs and DIAWs, and their MI
in a three-component DPM. Jukui and He \cite{Jukui2003} demonstrated the amplitude modulation of
spherical and cylindrical DIAWs. Saini and Kourakis \cite{Saini2008} considered a three-component
DPM having inertial highly charged massive dust grains and inertialess electrons and ions to study
the MI of DAWs, and highlighted that the angular frequency of the DAWs increases with the super-thermality
of the plasma species and the stable parametric regime decreases with an increase in the value of negative dust number density.

The outline of the paper is as follows: The governing equations describing our plasma
model are presented in section \ref{2sec:Governing Equations}. The derivation
of NLSE is demonstrated in section \ref{2sec:Derivation of NLSE}. The Modulational instability
and rogue waves is described in section \ref{2sec:Modulational instability and rogue waves}.
Results and discussion are devoted in section \ref{2sec:Results and discussion}. Conclusion is provided in section \ref{2sec:Conclusion}.
\section{Governing Equations}
\label{2sec:Governing Equations}
We consider an unmagnetized, fully ionized and collisionless three-component DPM comprising super-thermal
electrons, positively charged inertial warm ions and negatively charged dust grains.  At equilibrium, the
overall  charge neutrality conditions of our plasma system can be written as $n_{e0}+Z_dn_{d0}=Z_in_{i0}$,
where $n_{e0}$, $n_{d0}$, and $n_{i0}$ are the equilibrium electron, dust, and ion number
densities respectively, and $Z_d$ ($Z_i$) is the charge state of the negative (positive) dust grain (ion).
The normalized equations describing the system can be written as
\begin{eqnarray}
&&\hspace*{-1.3cm}\frac{\partial n_d}{\partial t}+\frac{\partial}{\partial x}(n_du_d)=0,
\label{2eq:1}\\
&&\hspace*{-1.3cm}\frac{\partial u_d}{\partial t}+u_d\frac{\partial u_d}{\partial x}=\mu_1\frac{\partial\phi}{\partial x},
\label{2eq:2}\\
&&\hspace*{-1.3cm}\frac{\partial n_i}{\partial t}+\frac{\partial}{\partial x}(n_iu_i)=0,
\label{2eq:3}\\
&&\hspace*{-1.3cm}\frac{\partial u_i}{\partial t}+u_i\frac{\partial u_i}{\partial x}+\mu_2n_i\frac{\partial n_i}{\partial x}=-\frac{\partial \phi}{\partial x},
\label{2eq:4}\\
&&\hspace*{-1.3cm}\frac{\partial^2\phi}{\partial x^2}+n_i=(1-\mu_3)n_e+\mu_3n_d,
\label{2eq:5}
\end{eqnarray}
where $n_d$ ($n_i$) is the dust (ion) number density normalized by the equilibrium value $n_{d0}$ ($n_{i0}$);
$u_d$ ($u_i$) is the dust (ion) fluid speed normalized by the ion sound speed $C_i=(Z_ik_BT_e/m_i)^{1/2}$
(where $T_e$ being the super-thermal electron temperature, $m_i$ is the ion rest mass, and $k_B$ is the Boltzmann
constant); $\phi$ is the electrostatic wave potential normalized by $k_BT_e/e$ (with e being the
electron charge). The time variable $t$ is normalized by $\omega_{P_i}^{-1}=[m_i/(4\pi e^2Z_i^2n_{i0})]^{1/2}$
and the space variable $x$ is normalized by $\lambda_{D_i}=(k_BT_e/4\pi e^2Z_in_{i0})^{1/2}$. The pressure of the
ion is expressed as $P_i=P_{i0}(N_i/N_{i0})^\gamma$, where $P_{i0}=n_{i0}k_BT_i$ being the equilibrium pressure of
 the ion, and $T_i$ being the warm ion temperature, and $\gamma=(N+2)/N$ (where $N$ be the degree
of freedom and for one dimensional case $N=1$, then $\gamma=3$).  Other plasma parameters are defined as
$\mu_1=\rho\mu$, $\rho=Z_d/Z_i$, $\mu=m_i/m_d$, $\mu_2=3T_i/Z_iT_e$, and $\mu_3=Z_dn_{d0}/Z_in_{i0}$.
The expression for the number density of the super-thermal electrons (following the
$\kappa$-distribution) can be expressed as
\begin{eqnarray}
&&\hspace*{-1.3cm}n_e=\Big[1-\frac{\phi}{\kappa-3/2}\Big]^{-\kappa+\frac{1}{2}}=1+ n_1\phi+ n_2\phi^2 +n_3\phi^3+ \cdot\cdot\cdot,
\label{2eq:6}\
\end{eqnarray}
where $n_1 =(2\kappa-1)/(2\kappa-3)$, $n_2 =[(2\kappa-1)(2\kappa+1)]/2(2\kappa-3)^2$, and $n_3 =[(2\kappa-1)(2\kappa+1)(2\kappa+3)]/6(2\kappa-3)^3$. The parameter $\kappa$, generally stands for super-thermality, which measures the deviation of the plasma particles from Maxwellian distribution.
Now, by substituting  Eq. \eqref{2eq:6} into \eqref{2eq:5} and expanding up to third order in $\phi$, we get
\begin{eqnarray}
&&\hspace*{-1.3cm}\frac{\partial^2\phi}{\partial x^2}+n_i+\mu_3=1+\mu_3n_d+A_1\phi+A_2\phi^2+A_3\phi^3+\cdot\cdot\cdot,
\label{2eq:7}
\end{eqnarray}
where $A_1=n_1(1-\mu_3)$, $A_2=n_2(1-\mu_3)$, and $A_3=n_3(1-\mu_3)$.
\section{Derivation of the NLSE}
\label{2sec:Derivation of NLSE}
To study the MI of the DIAWs, we want to derive the NLSE by employing the RPM.
First, we can write the stretched co-ordinates in the following form \cite{C6,C7,C8,C9,C10}
\begin{eqnarray}
&&\hspace*{-1.3cm}\xi=\epsilon(x-v_g t),
\label{2eq:8}\\
&&\hspace*{-1.3cm}\tau=\epsilon^2 t,
\label{2eq:9}
\end{eqnarray}
where $v_g$ is the group velocity and $\epsilon$ ($0<\epsilon<1$) is a small parameter.
We can write the dependent variables ($n_d$, $u_d$, $n_i$, $u_i$, and $\phi$) as
\begin{eqnarray}
&&\hspace*{-1.3cm}n_{d}=1+\sum_{m=1}^{\infty}\epsilon^{m}\sum_{l=-\infty}^{\infty}n_{dl}^{(m)}(\xi,\tau)~\mbox{e}^{i l(kx-\omega t)},
\label{2eq:10}\\
&&\hspace*{-1.3cm}u_{d}=\sum_{m=1}^{\infty}\epsilon^{m}\sum_{l=-\infty}^{\infty}u_{dl}^{(m)}(\xi,\tau)~\mbox{e}^{i l(kx-\omega t)},
\label{2eq:11}\\
&&\hspace*{-1.3cm}n_{i}=1+\sum_{m=1}^{\infty}\epsilon^{m}\sum_{l=-\infty}^{\infty}n_{il}^{(m)}(\xi,\tau)~\mbox{e}^{i l(kx-\omega t)},
\label{2eq:12}\\
&&\hspace*{-1.3cm}u_{i}=\sum_{m=1}^{\infty}\epsilon^{m}\sum_{l=-\infty}^{\infty}u_{il}^{(m)}(\xi,\tau)~\mbox{e}^{i l(kx-\omega t)},
\label{2eq:13}\\
&&\hspace*{-1.3cm}\phi=\sum_{m=1}^{\infty}\epsilon^{m}\sum_{l=-\infty}^{\infty}\phi_{l}^{(m)}(\xi,\tau)~\mbox{e}^{i l(kx-\omega t)},
\label{2eq:14}\
\end{eqnarray}
where $k$ and $\omega$ are the real variables representing the carrier wave number and
frequency, respectively. The derivative operators can be written as
\begin{eqnarray}
&&\hspace*{-1.3cm}\frac{\partial}{\partial x}\rightarrow\frac{\partial}{\partial x}+\epsilon\frac{\partial}{\partial\xi},
\label{2eq:15}\\
&&\hspace*{-1.3cm}\frac{\partial}{\partial t}\rightarrow\frac{\partial}{\partial t}-\epsilon v_g\frac{\partial}{\partial\xi}+\epsilon^2\frac{\partial}{\partial\tau}.
\label{2eq:16}
\end{eqnarray}
Now, by substituting Eqs, \eqref{2eq:8}-\eqref{2eq:16}  into  Eqs. \eqref{2eq:1}-\eqref{2eq:4}, and \eqref{2eq:7}, and
collecting the terms containing $\epsilon$, the first order ($m=1$ with $l=1$)  reduced equations can be written as
\begin{eqnarray}
&&\hspace*{-1.3cm}u_{d1}^{(1)}=-\frac{k\mu_1}{\omega}\phi_1^{(1)},
\label{2eq:17}\\
&&\hspace*{-1.3cm}n_{d1}^{(1)}=-\frac{\mu_1k^2}{\omega^2}\phi_1^{(1)},
\label{2eq:18}\\
&&\hspace*{-1.3cm}u_{i1}^{(1)}=-\frac{k\omega}{\mu_2k^2-\omega^2}\phi_1^{(1)},
\label{2eq:19}\\
&&\hspace*{-1.3cm}n_{i1}^{(1)}=-\frac{k^2}{\mu_2k^2-\omega^2}\phi_1^{(1)},
\label{2eq:20}\
\end{eqnarray}
these relations provide the dispersion relations of DIAWs. Now, the dispersion relations of DIAWs are
\begin{eqnarray}
&&\hspace*{-1.3cm}\omega^2\equiv\omega_f^2=\frac{k^2M+k^2\sqrt{M^2-4GH}}{2G},
\label{2eq:21}
\end{eqnarray}
and
\begin{eqnarray}
&&\hspace*{-1.3cm}\omega^2\equiv\omega_s^2=\frac{k^2M-k^2\sqrt{M^2-4GH}}{2G},
\label{2eq:22}
\end{eqnarray}
\begin{figure}[t!]
\centering
\includegraphics[width=80mm]{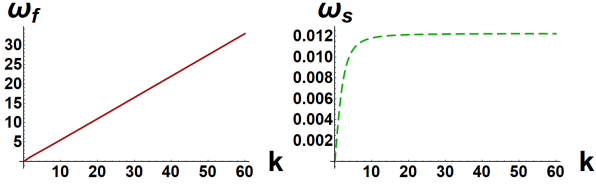}
\caption{The variation of $\omega_f$ vs $k$ (left panel) and $\omega_s$ vs $k$ (right panel)
when other plasma parameters are $\kappa=1.8$, $\rho=1\times10^3$,
$\mu=3\times10^{-6}$, $\mu_2=0.3$, and $\mu_3=0.05$.}
\label{2Fig:F1}
\end{figure}
where $M=1+\mu_1\mu_3+\mu_2k^2+A_1\mu_2$, $G=A_1+k^2$, and $H=\mu_1\mu_2\mu_3$. In Eqs. \eqref{2eq:21}
and \eqref{2eq:22}, to get the positive value of $\omega$, the condition $M^2>4GH$ must be satisfied.
In the fast $(\omega_f)$ DIA mode, both inertial ion and dust oscillate in phase with the
inertialess electrons. While in the slow $(\omega_s)$ DIA mode, one of the inertial elements ion (dust)
oscillates in phase with the inertialess electrons and other inertial element dust (ion) oscillates in
anti-phase with them \cite{Dubinov2009,Saberian2017}. Both the fast $(\omega_f)$ and
slow $(\omega_s)$ DIA modes have been analyzed numerically in Fig. \ref{2Fig:F1} in the presence of super-thermal electrons.
The second-order (when $m=2$ with $l=1$) equations are given by
\begin{eqnarray}
&&\hspace*{-1.3cm}u_{d1}^{(2)}=-\frac{k\mu_1}{\omega}\phi_1^{(2)}-\frac{\mu_1}{i\omega}\frac{\partial\phi_1^{(1)}}{\partial\xi}
+\frac{kv_g\mu_1}{i\omega^2}\frac{\partial\phi_1^{(1)}}{\partial\xi},
\label{2eq:23}\\
&&\hspace*{-1.3cm}n_{d1}^{(2)}=-\frac{\mu_1 k^2}{\omega^2}\phi_1^{(2)}+\frac{2k\mu_1(kv_g-\omega)}{i\omega^3}\frac{\partial\phi_1^{(1)}}{\partial\xi},
\label{2eq:24}\\
&&\hspace*{-1.3cm}u_{i1}^{(2)}=\frac{k\omega}{\omega^2-\mu_2k^2}\phi_1^{(2)}-\frac{i(\omega-kv_g)(\omega^2+\mu_2k^2)}{(\omega^2-\mu_2k^2)^2}\frac{\partial\phi_1^{(1)}}{\partial\xi},
\label{2eq:25}\\
&&\hspace*{-1.3cm}n_{i1}^{(2)}=\frac{k^2}{\omega^2-\mu_2k^2}\phi_1^{(2)}-\frac{2ik\omega(\omega-kv_g)}{(\omega^2-\mu_2k^2)^2}\frac{\partial\phi_1^{(1)}}{\partial\xi},
\label{2eq:26}
\end{eqnarray}
with the compatibility condition
\begin{eqnarray}
&&\hspace*{-1.3cm}v_g=\frac{2\mu_1\mu_3k\omega(\omega^2-\mu_2k^2)^2+2k\omega^5-2k\omega^3(\omega^2-\mu_2k^2)^2}{2k^2\mu_1\mu_3(\omega^2-\mu_2k^2)^2+2k^2\omega^4}.
\label{2eq:27}
\end{eqnarray}
The coefficients of the $\epsilon$ when $m=2$ with $l=2$ provides the second-order harmonic amplitudes
which are found to be proportional to $|\phi_1^{(1)}|^2$
\begin{eqnarray}
&&\hspace*{-1.3cm}u_{d2}^{(2)}=A_4|\phi_1^{(1)}|^2,
\label{2eq:28}\\
&&\hspace*{-1.3cm}n_{d2}^{(2)}=A_5|\phi_1^{(1)}|^2,
\label{2eq:29}\\
&&\hspace*{-1.3cm}u_{i2}^{(2)}=A_6|\phi_1^{(1)}|^2,
\label{2eq:30}\\
&&\hspace*{-1.3cm}n_{i2}^{(2)}=A_7|\phi_1^{(1)}|^2,
\label{2eq:31}\\
&&\hspace*{-1.3cm}\phi_{2}^{(2)}=A_8|\phi_1^{(1)}|^2,
\label{2eq:32}\
\end{eqnarray}
where
\begin{eqnarray}
&&\hspace*{-1.3cm}A_4=\frac{k^3\mu_1^2-2A_8\mu_1k\omega^2}{2\omega^3},
\nonumber\\
&&\hspace*{-1.3cm}A_5=\frac{3k^4\mu_1^2-2A_8\mu_1k^2\omega^2}{2\omega^4},
\nonumber\\
&&\hspace*{-1.3cm}A_6=-\frac{2A_8k\omega(\mu_2k^2-\omega^2)^2+3\mu_2\omega k^5+k^3\omega^3}{2(\mu_2k^2-\omega^2)^3},
\nonumber\\
&&\hspace*{-1.3cm}A_7=-\frac{2A_8k^2(\mu_2k^2-\omega^2)^2+3\omega^2k^4+\mu_2k^6}{2(\mu_2k^2-\omega^2)^3},
\nonumber\\
&&\hspace*{-1.3cm}A_8=\frac{(2A_2\omega^4+3\mu_1^2\mu_3k^4)+3\omega^6k^4+\mu_2k^6\omega^4}{2\omega^2
\big[(k^2\mu_1\mu_3-4k^2\omega^2-A_1\omega^2)-k^2\omega^2(\mu_2k^2-\omega^2)^{-1}\big]}.
\nonumber\
\end{eqnarray}
When $m=3$ with $l=0$ and $m=2$ with $l=0$ lead to zeroth harmonic modes as follows
\begin{eqnarray}
&&\hspace*{-1.3cm}u_{d0}^{(2)}=A_9|\phi_1^{(1)}|^2,
\label{2eq:33}\\
&&\hspace*{-1.3cm}n_{d0}^{(2)}=A_{10}|\phi_1^{(1)}|^2,
\label{2eq:34}\\
&&\hspace*{-1.3cm}u_{i0}^{(2)}=A_{11}|\phi_1^{(1)}|^2,
\label{2eq:35}\\
&&\hspace*{-1.3cm}n_{i0}^{(2)}=A_{12}|\phi_1^{(1)}|^2,
\label{2eq:36}\\
&&\hspace*{-1.3cm}\phi_0^{(2)}=A_{13} |\phi_1^{(1)}|^2,
\label{2eq:37}\
\end{eqnarray}
where
\begin{eqnarray}
&&\hspace*{-1.3cm}A_9=\frac{k^2\mu_1^2-\mu_1\omega^2A_{13}}{\omega^2v_g},
\nonumber\\
&&\hspace*{-1.3cm}A_{10}=\frac{2k^3\mu_1^2v_g+k^2\mu_1^2\omega-\mu_1\omega^3A_{13}}{v_g^2\omega^3},
\nonumber\\
&&\hspace*{-1.3cm}A_{11}=\frac{v_g^2A_{13}(\omega^2-\mu_2k^2)^2+2\mu_2\omega k^3+\mu_2k^4v_g+k^2\omega^2v_g}{(\omega^2-\mu_2k^2)^2(v_g^2-\mu_2)},
\nonumber\\
&&\hspace*{-1.3cm}A_{12}=\frac{A_{13}(\omega^2-\mu_2k^2)^2+2\omega k^3v_g+\mu_2k^4+k^2\omega^2}{(\omega^2-\mu_2k^2)^2(v_g^2-\mu_2)},
\nonumber\\
&&\hspace*{-1.3cm}A_{13}=\frac{(\omega^2-\mu_2k^2)^2\times F_1-v_g^2\omega^3(2\omega k^3v_g+\mu_2k^4+k^2\omega^2)}
{\omega^3(\omega^2-\mu_2k^2)^2\big[\mu_1\mu_3(v_g^2-\mu_2)+v_g^2-A_1v_g^2(v_g^2-\mu_2)\big]},
\nonumber\
\end{eqnarray}
where $F_1=(v_g^2-\mu_2)(2A_2v_g^2\omega^3+2k^3\mu_1^2\mu_3v_g+k^2\omega\mu_1^2\mu_3)$.
Finally, the third-order harmonic modes (when $m=3$ and $l=1$) and with the help of Eqs. \eqref{2eq:17}-\eqref{2eq:37},
given a set of equations which can be reduced to the standard NLSE:
\begin{eqnarray}
&&\hspace*{-1.3cm}i\frac{\partial\Phi}{\partial\tau}+P\frac{\partial^2\Phi}{\partial\xi^2}+Q|\Phi|^2\Phi=0,
\label{2eq:38}
\end{eqnarray}
where $\Phi=\phi_1^{(1)}$ for simplicity. In Eq. \eqref{2eq:38}, the dispersion
coefficients ($P$) and non-linear coefficients ($Q$) can be written, respectively, as
\begin{eqnarray}
&&\hspace*{-1.3cm}P=-\frac{(\omega-kv_g)(\omega^2-\mu_2k^2)^3(3kv_g\mu_1\mu_3-\mu_1\mu_2\omega)+F_2}
{\omega(\omega^2-\mu_2k^2)\big[2\mu_1\mu_3k^2(\omega^2-\mu_2k^2)^2+2k^2\omega^4\big]},
\nonumber\\
&&\hspace*{-1.3cm}Q=\frac{\omega^3(\omega^2-\mu_2k^2)^2[3A_3+2A_2(A_8+A_{13})-F_3]}{2\mu_1\mu_3k^2(\omega^2-\mu_2k^2)^2+2k^2\omega^4},
\nonumber\
\end{eqnarray}
where $F_2=(2k\omega^6v_g-2\mu_2k^2\omega^5)(\omega-kv_g)+(\omega^4kv_g-\omega^5)(\omega-kv_g)(\omega^2+\mu_2k^2)+\omega^4(\omega^2-\mu_2k^2)^3$ and $F_3=\{(k^2\omega^2+\mu_2k^4)(A_7+A_{12})+2k^3\omega(A_6+A_{11})\}/(\omega^2-\mu_2k^2)^2 + \{2k^3\mu_1\mu_3(A_4+A_9)+k^2\omega\mu_1\mu_3(A_5+A_{10})\}/\omega^3$. It is
interesting that $P$ and  $Q$ of the Eq. \eqref{2eq:38} are function of various plasma parameters such as carrier wave number ($k$),
ratio of ion mass to dust mass ($\mu$), ratio of dust charge state to ion charge state ($\rho$), and super-thermal parameter ($\kappa$), etc.
\begin{figure}[t!]
\centering
\includegraphics[width=80mm]{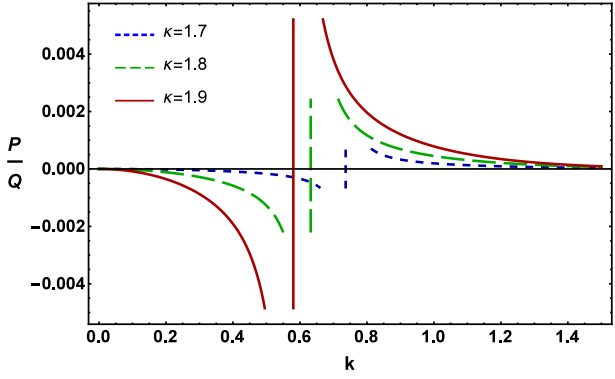}
\caption{The variation of $P/Q$ with $k$ for different values of $\kappa$ when other plasma
parameters are $\rho=1\times10^3$, $\mu=3\times10^{-6}$, $\mu_2=0.3$, $\mu_3=0.05$, and $\omega\equiv\omega_f$.}
\label{2Fig:F2}
\end{figure}
\begin{figure}[t!]
\centering
\includegraphics[width=80mm]{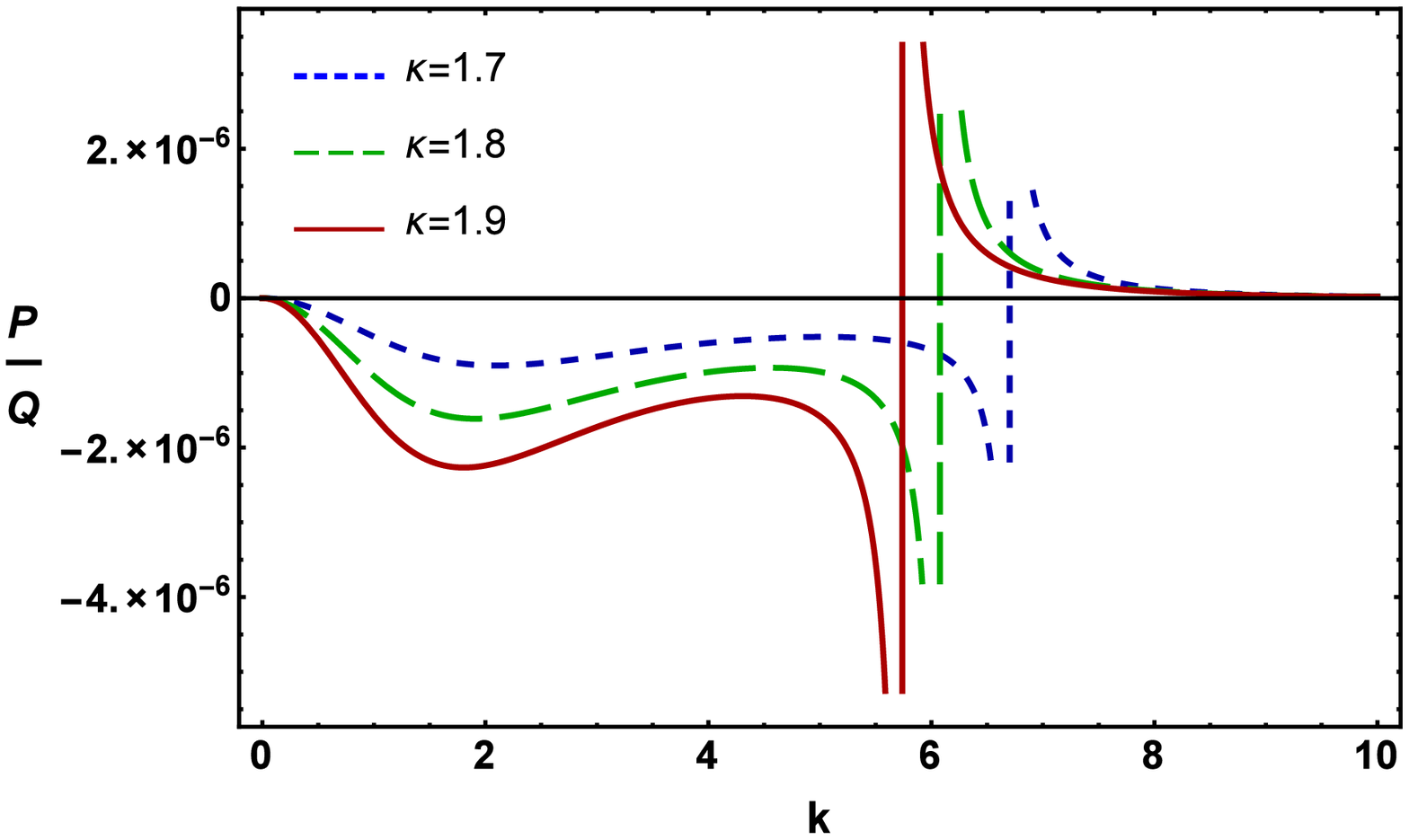}
\caption{The variation of $P/Q$ with $k$ for different values of $\kappa$ when other plasma
parameters are $\rho=1\times10^3$, $\mu=3\times10^{-6}$, $\mu_2=0.3$, $\mu_3=0.05$, and $\omega\equiv\omega_s$.}
 \label{2Fig:F3}
\end{figure}
\begin{figure}[t!]
\centering
\includegraphics[width=80mm]{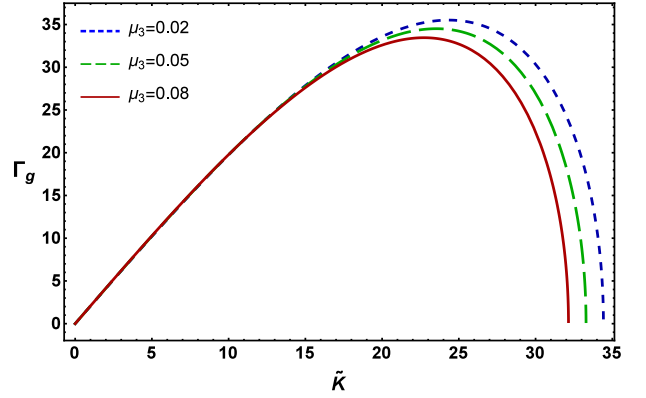}
\caption{The variation of $\Gamma_g$ with $\tilde{k}$ for different values of $\mu_3$ when other
plasma parameters are $k=1.0$, $\tilde{\Phi}_0=0.5$, $\kappa=1.8$, $\rho=1\times10^3$,
$\mu=3\times10^{-6}$, $\mu_2=0.3$, and $\omega\equiv\omega_f$.}
\label{2Fig:F4}
\end{figure}
\begin{figure}[t!]
\centering
\includegraphics[width=80mm]{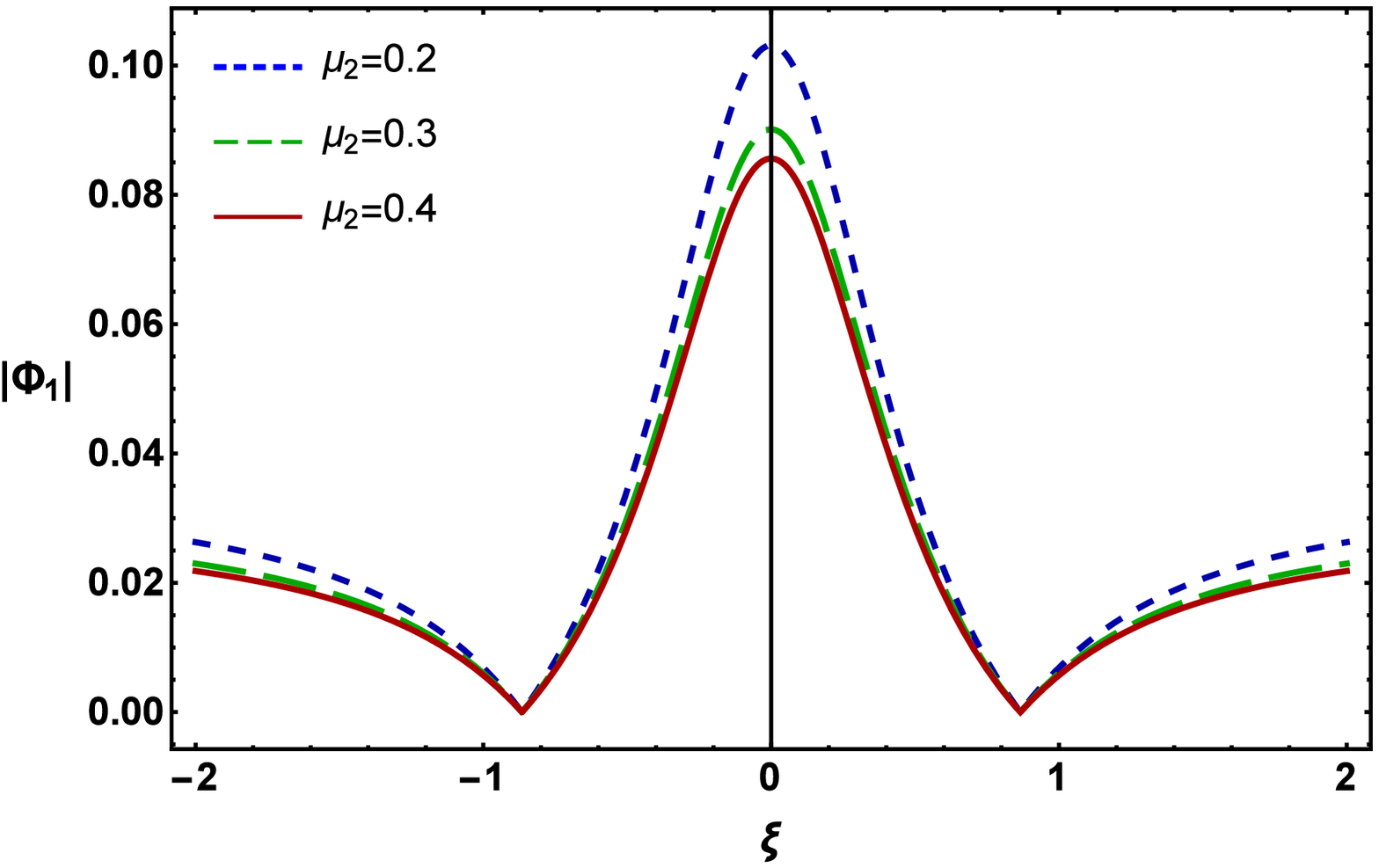}
\caption{The variation of $|\Phi_1|$ with $\xi$ for different values of $\mu_2$
when other plasma parameters are $\tau=0$, $k=1.0$, $\tilde{\Phi}_0=0.5$, $\kappa=1.8$, $\rho=1\times10^3$,
$\mu=3\times10^{-6}$, $\mu_3=0.05$, and $\omega\equiv\omega_f$.}
\label{2Fig:F5}
\end{figure}
\begin{figure}[t!]
\centering
\includegraphics[width=80mm]{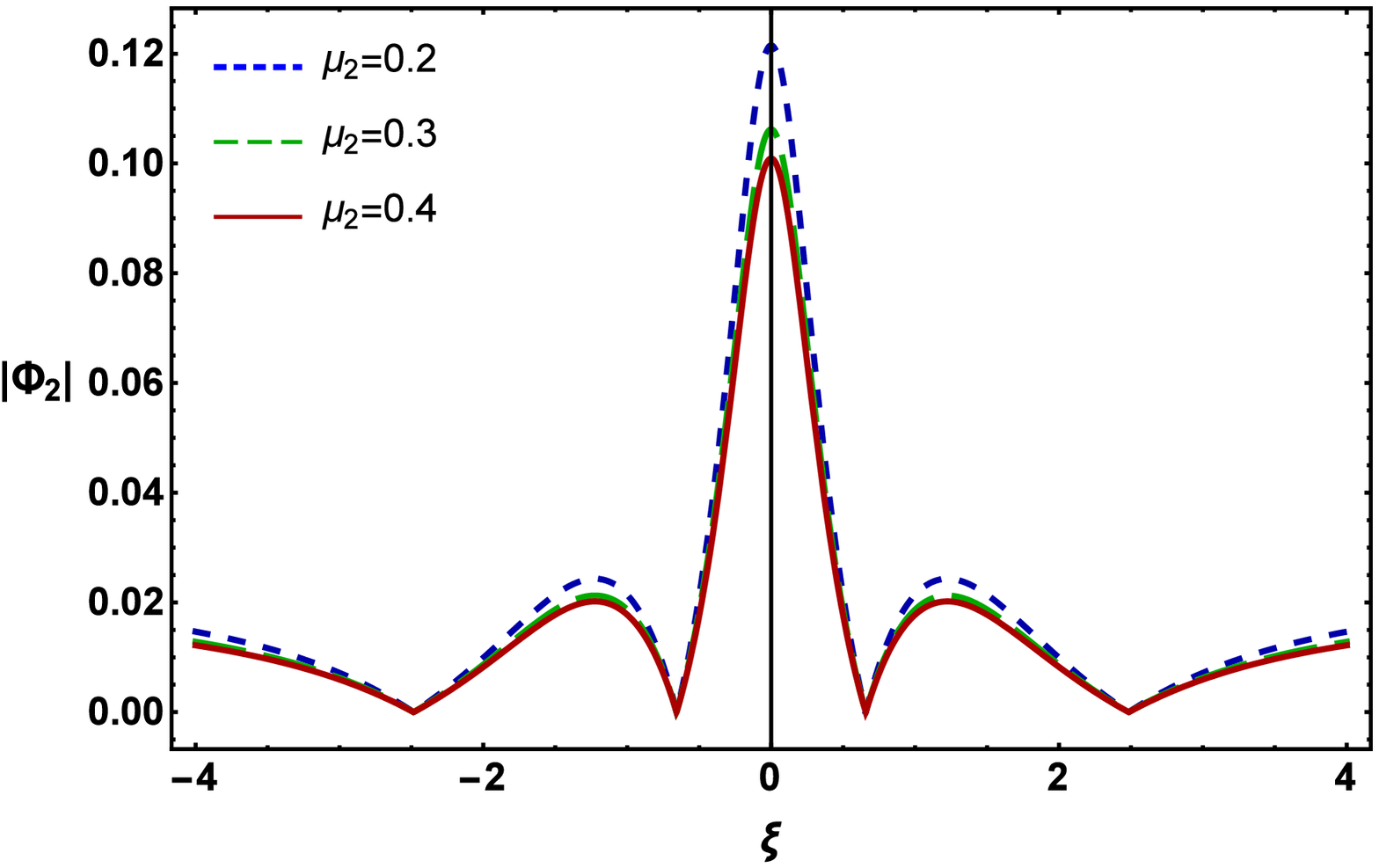}
\caption{The variation of $|\Phi_2|$ with $\xi$ for different values of $\mu_2$
when other plasma parameters are $\tau=0$, $k=1.0$, $\tilde{\Phi}_0=0.5$, $\kappa=1.8$, $\rho=1\times10^3$,
$\mu=3\times10^{-6}$, $\mu_3=0.05$, and $\omega\equiv\omega_f$.}
 \label{2Fig:F6}
\end{figure}
\begin{figure}[t!]
\centering
\includegraphics[width=80mm]{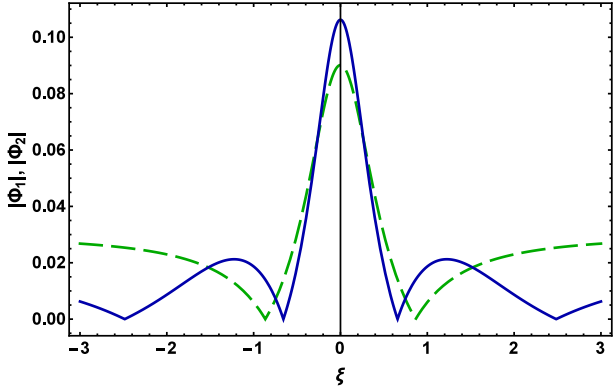}
\caption{The variation of first-order (dashed green curve) and
second-order (solid blue curve) rational solutions of NLSE at
$k=1.0$ and $\tau=0$.}
\label{2Fig:F7}
\end{figure}
\section{Modulational instability and rogue waves}
\label{2sec:Modulational instability and rogue waves}
The stable and unstable parametric regimes of the DIAWs are organized by the sign of the dispersion ($P$)
and nonlinear ($Q$) coefficients of the standard NLSE \cite{Kourakis2003,Kourakis2005,Chowdhury2018}.
The stability of DIAWs in a three-component DPM is governed by the sign of $P$
and $Q$ \cite{Kourakis2003,Kourakis2005,Chowdhury2018}. When $P$ and $Q$ have same sign (i.e., $P/Q>0$),
the evolution of the DIAWs amplitude is modulationally unstable. On the other hand, when $P$ and $Q$ have
opposite sign (i.e., $P/Q<0$), the DIAWs are modulationally stable in the presence of external perturbations.
The plot of $P/Q$ against $k$ yields stable and unstable parametric regimes of DIAWs.
The point, at which transition of $P/Q$ curve intersects with $k$-axis, is known as threshold
or critical wave number $k$ ($=k_c$) \cite{Kourakis2003,Kourakis2005,Chowdhury2018,C11,C12,C13}.
When $P/Q>0$ and $\tilde{k}<k_c$, the MI growth rate $(\Gamma_g)$ is given by
\begin{eqnarray}
&&\hspace*{-1.3cm}\Gamma_g=|P|\tilde{k}^2\sqrt{\frac{\tilde{k}_c^2}{\tilde{k}^2}-1}.
\label{2eq:39}
\end{eqnarray}
The first-order rational solution, which can predict the concentration of the large amount of energy
in a small region in the modulationally unstable parametric regime ($P/Q>0$)
of DIAWs, of Eq.  \eqref{2eq:38} can be written as \cite{Ankiewiez2009,Guo2012}
\begin{eqnarray}
&&\hspace*{-1.3cm}\Phi_1 (\xi, \tau)=\sqrt{\frac{2P}{Q}}\Big[\frac{4+16 i\tau P}{1+4 \xi^2 + 16\tau^2 P^2}-1\Big] \mbox{exp} (2i\tau P).
\label{2eq:40}
\end{eqnarray}
and the second-order rational solution is
\begin{eqnarray}
&&\hspace*{-1.3cm}\Phi_2 (\xi, \tau)=\sqrt{\frac{P}{Q}}\Big[1+\frac{G_2(\xi,\tau)+iM_2(\xi,\tau)}{D_2(\xi,\tau)}\Big]\mbox{exp}(i\tau P),
\label{2eq:41}\
\end{eqnarray}
where
\begin{eqnarray}
&&\hspace*{-1.3cm}G_2(\xi,\tau)=\frac{-\xi^4}{2}-6(P\xi\tau)^2-10(P\tau)^4
\nonumber\\
&&\hspace*{0.4cm}-\frac{3\xi^2}{2}-9(P\tau)^2+\frac{3}{8},
\nonumber\\
&&\hspace*{-1.3cm}M_2(\xi,\tau)=-P\tau\Big[\xi^4+4(P\xi\tau)^2+4(P\tau)^4
\nonumber\\
&&\hspace*{0.4cm}-3\xi^2+2(P\tau)^2-\frac{15}{4}\Big],
\nonumber\\
&&\hspace*{-1.3cm}D_2(\xi,\tau)=\frac{\xi^6}{12}+\frac{\xi^4(P\tau)^2}{2}+\xi^2(P\tau)^4
\nonumber\\
&&\hspace*{0.4cm}+\frac{\xi^4}{8}+\frac{9(P\tau)^4}{2}-\frac{3(P\xi\tau)^2}{2}
\nonumber\\
&&\hspace*{0.4cm}+\frac{9\xi^2}{16}+\frac{33(P\tau)^2}{8}+\frac{3}{32}.
\nonumber\
\end{eqnarray}
The nonlinear behavior of the plasma medium is considered to be responsible for the
concentration of large amount of energy into tiny region.
\section{Results and discussion}
\label{2sec:Results and discussion}
First, we are interested to observe numerically the stable and unstable parametric regimes
of DIAWs in the presence of super-thermal electrons by depicting the
variation of $P/Q$ with $k$ for different plasma parameters. In our present analysis,
we have considered that $m_d=10^6m_i$, $Z_d=10^3Z_i$, and $T_e=10T_i$.

We have graphically shown the variation of $P/Q$ with $k$ in case of both fast ($\omega_f$)
and slow ($\omega_s$) DIA modes for different values of $\kappa$ in
Figs. \ref{2Fig:F2} and \ref{2Fig:F3}, respectively.  From these two figures, it can be seen
that (a) under consideration $\omega\equiv\omega_f$ and $\omega\equiv\omega_s$, possible stable
and unstable parametric regimes can be occurred for DIAWs; (b) the DIAWs become unstable for small value of
$k$ (i.e., $k\simeq0.6$) in first mode while in slow mode the DIAWs become unstable for large value
of $k$ (i.e., $k\simeq6$) for same plasma parameters; and (c)  the $k_c$ decreases with an increase in the value of $\kappa$.

We have numerically analyzed the MI growth rate of DIAWs under consideration fast mode in Fig. \ref{2Fig:F4}
by using these plasma parameters: $k=1.0$, $\tilde{\Phi}_0=0.5$, $\rho=1\times10^3$, $\mu=3\times10^{-6}$, and
$\mu_2=0.3$. It is clear that the maximum value of the $\Gamma_g$ increases (decreases) with the increase in the value of
ion (dust) number density for a constant value of their charge state (via $\mu_3$).
The nonlinearity as well as the $\Gamma_g$ increases (decreases) with ion (dust) charge state when other plasma parameters
remain constant.

We have presented the evaluation of first and second-order DIARWs with $\xi$ for different
values of $\mu_2$ in Figs. \ref{2Fig:F5} and \ref{2Fig:F6}, respectively, and from these figures it
is observed that both the first and second-order DIARW solutions can concentrate large
amount of energy into a small region. It is clear from these two figures that (a)  the amplitude of
the first and second-order rogue waves decreases (increases) with increasing the temperature of the
ion (electron) for a fixed ion charge state; (b) the nonlinearity
as well as the amplitude and width of the first and second-order DIARWs increases with ion charge state.

Figure \ref{2Fig:F7} shows a comparison between the first and second-order DIARWs and it can be seen from this
figure that (a) the amplitude of the second-order DIARWs is always higher than the first-order DIARWs
for same plasma parameters, which means that the second-order DIARWs can concentrate more energy
than the first-order DIARWs; (b) the first-order DIARWs has two zeros symmetrically located on
its $\xi$-axis, where the second-order DIARWs has four zeros symmetrically located on its $\xi$-axis.
\section{Conclusion}
\label{2sec:Conclusion}
In this paper, we have considered a realistic DPM having negatively charged dust grains,
ions and electrons. A standard NLSE is derived by using RPM, and this three-component DPM
can generate DIAWs in which the moment of inertia is provided by the warm ions and dust
grains, and the restoring force is provided by the thermal pressure of inertialess
super-thermal electrons. The interaction of the nonlinear ($Q$) and dispersive ($P$)
coefficients of NLSE can easily divide the modulationally stable and unstable parametric
regimes, and the unstable parametric regime also allows  to generate highly energetic DIARWs.
The outcomes of present investigation can be useful in explaining the DIARWs  cometary
tails \cite{Eslami2013,Saini2016,Shahmansouri2012,Shukla2002},  magnetosphere \cite{Shahmansouri2012},
ionosphere \cite{Shahmansouri2012}, aerosols in the astrosphere \cite{Saini2016},
planetary rings \cite{Eslami2013}, Earth's ionospheres \cite{Eslami2013}, and interstellar medium \cite{Shukla2002}.

\end{document}